\documentclass[12pt,a4paper]{article}
\usepackage[utf8]{inputenc}
\usepackage[english]{babel}
\usepackage{amsmath, amsfonts, amssymb, amsthm}
\usepackage{geometry}
\usepackage{graphicx}
\usepackage{hyperref}
\usepackage{listings}
\usepackage{xcolor}
\usepackage{bm}
\usepackage{cite}
\usepackage{setspace}
\usepackage{helvet}

\title{Pricing American Commodity Options Under the Schwartz Two-Factor Model Using the Alternating Direction Explicit (ADE) Method}
\author{Sanjith Prabhu \\ Dr. Daniel J. Duffy}
\date{\today}

\begin{document}

\maketitle
\newpage
\begin{abstract} \normalsize
This project investigates the valuation of American commodity options using the Schwartz two-factor model. We employ the double-sweep Alternating Direction Explicit (ADE) method, an unconditionally stable (for diffusion) finite difference scheme that provides the efficiency of explicit marching with the stability typically reserved for implicit methods, and compare it with the single-sweep alternating variant of Saul'yev. We derive the governing partial differential equation (PDE) and transform the semi-infinite domain into a rectangular computational domain. Furthermore, we apply Fichera theory to determine boundary conditions at the limits of the spot price. Finally, we implement the solver in C++, validating the results first against the closed-form solution of the PDE and Monte Carlo as a European option, then apply the American early exercise constraint and benchmark against Longstaff-Schwartz Monte Carlo to analyze numerical accuracy and convergence. The framework is then extended, following feedback from a commodities trader to exercise on futures prices and use seasonal parameters.
\end{abstract}

\newpage
\tableofcontents

\newpage
\pagenumbering{arabic}

\section{Introduction}\label{ch:intro}
Partial differential equations (PDEs) have been the basis of modern computational finance since the work of Black and Scholes (1973) and Merton (1973), who demonstrated that the value of a derivative security could be represented as the solution to a parabolic partial differential equation. Through It\^o calculus, mathematical techniques from numerical analysis have been applied directly to financial engineering. Although closed-form analytical solutions exist for a limited class of derivative contracts, most practical pricing problems involve nonlinear payoffs, multiple stochastic risk factors, or embedded contractual features that preclude analytical treatment. Consequently, numerical methods for solving financial PDEs have become indispensable tools within both academia and industry.\\

Among the various numerical approaches available, finite difference methods (FDMs) remain one of the most widely employed techniques for approximating solutions to financial pricing equations. By discretizing the spatial and temporal variables, finite difference methods replace continuous differential operators with algebraic approximations defined on a computational grid. Classical explicit, implicit and Crank--Nicolson schemes (Crank and Nicolson, 1947) have each become standard methodologies for option pricing, with different trade-offs between computational efficiency, numerical stability and convergence. More advanced operator splitting techniques extend these methods to multidimensional problems by decomposing complex differential operators into simpler components that may be solved sequentially. In particular, the Alternating Direction Explicit (ADE) method, which was first introduced by Saul'yev (1964), provides an efficient framework for solving multidimensional parabolic PDEs while preserving much of the simplicity of explicit finite difference schemes. Its accuracy and stability properties, relative to implicit schemes such as Crank--Nicolson, have been discussed in detail by Duffy (2009). A broad treatment of PDE and finite difference methods in computational finance is given in Duffy (2022).\\\\
In this project, we focus on American-style options, which present an additional numerical challenge because they allow exercise at any time prior to maturity. Unlike European contracts, whose values depend solely upon the terminal payoff, American options require determination of an optimal exercise strategy throughout the contract's lifetime. The resulting valuation problem becomes a free-boundary problem, in which the numerical solution must satisfy both the governing pricing equation and an inequality constraint ensuring that the option value never falls below its intrinsic value. Accurate approximation of this moving exercise boundary represents one of the principal computational challenges in American option pricing and motivates the development of robust numerical algorithms.\\\\
While the original Black--Scholes framework was developed for equity derivatives, subsequent research extended option pricing theory to futures and commodity markets. Black (1976) introduced a pricing framework for options written on futures contracts, providing one of the earliest theoretical foundations for commodity derivative valuation under no-arbitrage principles. However, commodity prices exhibit characteristics that differ fundamentally from those of financial assets. Physical commodities are influenced by storage costs, transportation costs, inventory levels and the economic benefits associated with holding the underlying asset as described by the theory of storage (Working, 1949 and Brennan, 1958). These benefits are commonly represented through the convenience yield, which reflects the implicit return obtained from maintaining physical inventories and plays a significant role in determining futures prices and term structures.\\\\
Recognizing these additional market characteristics, Gibson and Schwartz (1990) proposed a stochastic two-factor commodity pricing model in which the spot price and convenience yield evolve jointly as correlated stochastic processes. Their work demonstrated that allowing the convenience yield to vary stochastically significantly improved the ability of the model to reproduce observed commodity price dynamics compared to single-factor models. Schwartz (1997) subsequently generalized this framework by modeling the convenience yield as a mean-reverting Ornstein--Uhlenbeck process under the risk-neutral measure while retaining stochastic spot price dynamics. The resulting two-factor model possesses sufficient flexibility to generate a wide variety of observed futures term structures while remaining mathematically tractable, and it has since become one of the most influential models for commodity derivative valuation.\\\\
Despite the tractability of European contracts under the Schwartz framework, American-style commodity options remain considerably more difficult to value. The stochastic convenience yield introduces a second state variable, increasing the dimensionality of the governing pricing equation, while the early exercise feature introduces a moving free boundary for which no closed-form solution exists. Therefore, numerical methods become essential. FDMs approximate the solution of the governing PDE directly, whereas simulation-based methods estimate option values through the generation of stochastic sample paths. Among the latter, the Least Squares Monte Carlo (LSMC) method introduced by Longstaff and Schwartz (2001) has become one of the most widely adopted approaches for pricing American-style derivatives by approximating continuation values through least-squares regression.\\\\
This project investigates the numerical pricing of American crude oil put options under the two-factor Schwartz (1997) commodity pricing model. The governing two-dimensional pricing equation is derived in Chapter~\ref{ch:model} in terms of the spot price and convenience yield, and is transformed to the logarithm of the spot price in Chapter~\ref{ch:apply}. Chapters~\ref{ch:ade} and~\ref{ch:apply} introduce the double-sweep ADE scheme, together with the original Saul'yev alternating variant as a comparisn, and apply them to the transformed PDE, decomposing the multidimensional operator into directional sweeps that retain the computational simplicity of explicit methods (Duffy, 2006). The American early exercise constraint is enforced after each backward time step (Chapter~\ref{ch:american}). In Chapter~\ref{ch:validation} we describe the C++ implementation, following the object-oriented framework of Duffy (2004), together with the validation strategy: the solver without early exercise is compared with the closed-form European price and Monte Carlo, and the American prices are benchmarked against Longstaff--Schwartz LSMC under identical parameters, with grid refinement studies and a comparison of the two ADE variants in Chapter~\ref{ch:results}. Chapter~\ref{ch:further} reports extensions suggested by a commodities trader: exercise on a futures contract, seasonal parameters, and two-factor Greeks with hedge ratios.\\\\
Rather than proposing a new stochastic model, the primary contribution of this project lies in the numerical implementation of ADE within the Schwartz commodity pricing framework. By examining the performance of the ADE finite difference solver against a widely accepted Monte Carlo benchmark, this study demonstrates the effectiveness of ADE methods for solving two-dimensional American commodity option pricing problems while providing further insight into the practical implementation of finite difference methods in computational finance.

\newpage

\section{The Two-Factor Commodity Model}\label{ch:model}

\subsection{Background on the Schwartz Two-Factor Model}
The stochastic behavior of commodity prices is fundamentally distinct from that of financial assets due to the presence of mean-reverting dynamics and the physical benefits associated with immediate availability. Following the framework originally proposed by Gibson and Schwartz (1990) and further refined by Schwartz (1997), we utilize a two-factor structural model to capture these complexities. This framework assumes that the commodity is fully described by its spot price and its instantaneous convenience yield. By treating the convenience yield as a stochastic variable, the model replicates the term structures observed in commodity markets that standard one-factor models fail to address.\\\\
It is worth noting that a third factor, stochastic interest rates, has been explored in both Miltersen and Schwartz (1998) and Hilliard and Reis (1998). However, the inclusion of this factor was not found to have a significant impact on the pricing of commodity options, motivating the use of the two-factor model in this project.\\\\
To estimate model parameters, Schwartz (1997) applies the Kalman filter method, specifically testing on copper, oil, and gold. The mathematical background of Kalman filtering for single-factor and multi-factor Schwartz models is extensively discussed in Rigatos et al. (2018). In Chapter~\ref{ch:results} of this project, we use the WTI crude oil parameters estimated by Schwartz (1997) for our numerical testing.

\subsection{Joint Stochastic Process}\label{sec:joint}

Following the framework introduced by Gibson and Schwartz (1990), further developed by Schwartz (1997), and the notation used by Rigatos et al. (2018), the commodity market is represented by two state variables: the spot price of the commodity $S_t$ and the instantaneous convenience yield $\delta_t$. The convenience yield represents the economic benefit obtained from physically holding the commodity rather than holding a financial claim on the commodity. This factor captures the effects of inventory scarcity, storage benefits, and other market conditions that influence the relationship between spot and futures prices. The importance of incorporating additional fundamental factors into commodity pricing models is supported by empirical observations that traditional storage models with constant fundamental volatility are unable to fully explain certain commodity market dynamics, including episodes where inventories and prices move together and the behavior of forward price variances and option-implied volatilities (Pirrong, 2012).\\\\
Under the physical measure, the spot price is assumed to follow a geometric diffusion whose expected return is reduced by the convenience yield, while the convenience yield follows a mean-reverting Ornstein--Uhlenbeck process. The joint dynamics of the two factors are given by
\begin{align}
    dS_t &= (\mu-\delta_t)S_t\,dt+\sigma_1S_t\,dz_{1,t}\label{eq:S_real}\\
    d\delta_t &= \kappa(\alpha-\delta_t)\,dt+\sigma_2\,dz_{2,t}\label{eq:d_real}
\end{align}
where $dz_{1,t}$ and $dz_{2,t}$ are increments of standard Brownian motions satisfying the correlation condition
\begin{equation}
    dz_{1,t}\,dz_{2,t}=\rho\,dt\label{eq:corr}
\end{equation}
The parameter $\mu$ represents the expected rate of return on the commodity, while $\sigma_1$ and $\sigma_2$ denote the volatilities of the spot price and convenience yield processes, respectively. The parameter $\kappa>0$ controls the speed at which the convenience yield reverts toward its long-run equilibrium level $\alpha$. The Ornstein--Uhlenbeck specification implies that temporary deviations in the convenience yield caused by inventory shocks decay over time, with stationary distribution
\begin{equation}
    \delta_\infty\sim\mathcal{N}\!\left(\alpha,\ \frac{\sigma_2^2}{2\kappa}\right),
\end{equation}
which will later guide the truncation of the computational domain in the $\delta$ direction.\\\\
The correlation parameter $\rho$ is important because periods of commodity scarcity may simultaneously increase both the value of immediate ownership and the spot price of the underlying commodity. The inclusion of this stochastic convenience yield and its correlation with the spot price represents one of the primary innovations of the Gibson and Schwartz (1990) framework, allowing the model to reproduce realistic commodity behavior.

\subsection{Adjusting Factors for Risk}

To value contingent claims written on the commodity, the stochastic processes must be expressed in a risk-adjusted form. Following Schwartz (1997), the spot commodity is treated as a traded asset whose expected return under risk-neutral valuation is equal to the risk-free rate adjusted for the convenience yield. Therefore, the drift of the spot price process changes from $(\mu-\delta)$ to $(r-\delta)$.\\\\
The convenience yield, however, represents an unobservable variable associated with holding physical inventory and cannot actually be traded. Schwartz (1997) incorporates this additional source of risk through a constant market price of convenience yield risk, denoted by $\lambda$. After adjusting for risk, the joint dynamics are
\begin{align}
    dS_t &= (r-\delta_t)S_t\,dt+\sigma_1S_t\,dz_{1,t}\label{eq:S_rn}\\
    d\delta_t &= \left[\kappa(\alpha-\delta_t)-\lambda\right]dt+\sigma_2\,dz_{2,t}\label{eq:d_rn}
\end{align}
with the same correlation as before in \eqref{eq:corr}. A positive value of $\lambda$ reduces the risk-adjusted equilibrium level of the convenience yield, reflecting the effect of the convenience yield risk premium.\\\\
For numerical implementation, it is convenient to absorb the market price of risk directly into the mean-reversion level. Defining the risk-adjusted long-run convenience yield as
\begin{equation}
    \hat{\alpha}=\alpha-\frac{\lambda}{\kappa},
\end{equation}
the convenience yield process becomes
\begin{equation}
    d\delta_t=\kappa(\hat{\alpha}-\delta_t)\,dt+\sigma_2\,dz_{2,t}\label{eq:d_rn2}
\end{equation}
This transformation preserves the original mean-reverting structure of the convenience yield while incorporating the risk adjustment required for derivative valuation. The resulting processes form the basis for the two-factor pricing equation used in the numerical implementation.

\subsection{Derivation of the Governing Two-Factor PDE}\label{sec:pde}
Consider a derivative with value $V(S,\delta,t)$ written on the commodity. Following Schwartz (1997), the governing pricing equation is obtained by applying It\^o's lemma to the risk-adjusted state variables and imposing the no-arbitrage condition for the derivative value. Because the spot price and convenience yield are driven by correlated stochastic processes, the resulting differential contains both second-order diffusion terms and a cross-derivative term representing the covariance between the two factors.

Applying It\^o's lemma to $V(S_t,\delta_t,t)$ gives
\begin{align}
dV ={}& 
\frac{\partial V}{\partial t}dt
+\frac{\partial V}{\partial S}dS
+\frac{\partial V}{\partial \delta}d\delta
+\frac{1}{2}\frac{\partial^2 V}{\partial S^2}(dS)^2 +\frac{1}{2}\frac{\partial^2 V}{\partial \delta^2}(d\delta)^2
+\frac{\partial^2 V}{\partial S\partial\delta}dS\,d\delta
\end{align}
Using the quadratic variation relationships
\begin{equation}
    (dS)^2=\sigma_1^2S^2dt,
    \qquad
    (d\delta)^2=\sigma_2^2dt,
    \qquad
    dS\,d\delta=\rho\sigma_1\sigma_2S\,dt,
\end{equation}
and substituting the risk-adjusted dynamics from \eqref{eq:S_rn} and \eqref{eq:d_rn2} gives
\begin{align}
dV ={}&
V_tdt
+V_S\left[(r-\delta)Sdt+\sigma_1Sdz_1\right] + V_\delta\left[\kappa(\hat{\alpha}-\delta)dt+\sigma_2dz_2\right] \notag\\
&+\frac{1}{2}V_{SS}\sigma_1^2S^2dt
+\frac{1}{2}V_{\delta\delta}\sigma_2^2dt
+V_{S\delta}\rho\sigma_1\sigma_2S\,dt
\end{align}
Collecting the drift terms and the stochastic terms separately, this becomes
\begin{align}
dV ={}&
\left[
V_t
+(r-\delta)SV_S
+\kappa(\hat{\alpha}-\delta)V_\delta
+\frac{1}{2}\sigma_1^2S^2V_{SS}
+\rho\sigma_1\sigma_2SV_{S\delta}
+\frac{1}{2}\sigma_2^2V_{\delta\delta}
\right]dt\notag\\
&+\sigma_1SV_Sdz_1+\sigma_2V_\delta dz_2
\end{align}
Under risk-neutral valuation, the expected return of the derivative must equal the risk-free rate. Therefore, the drift component of the derivative value must satisfy ${E}[dV]=rV\,dt$. Equating the drift term to the risk-free return gives the governing two-factor pricing equation
\begin{equation}
\frac{1}{2}\sigma_1^2S^2V_{SS}
+
\rho\sigma_1\sigma_2SV_{S\delta}
+
\frac{1}{2}\sigma_2^2V_{\delta\delta}
+
(r-\delta)SV_S
+
\kappa(\hat{\alpha}-\delta)V_\delta
+
V_t
-rV=0.
\end{equation}
Introducing the time-to-maturity variable $\tau=T-t$ gives the backward-time form used for numerical solution,
\begin{equation}\label{eq:pde_tau}
V_\tau=\frac{1}{2}\sigma_1^2S^2V_{SS}
+
\rho\sigma_1\sigma_2SV_{S\delta}
+
\frac{1}{2}\sigma_2^2V_{\delta\delta}
+
(r-\delta)SV_S
+
\kappa(\hat{\alpha}-\delta)V_\delta
-rV
\end{equation}
The cross-derivative term $\rho\sigma_1\sigma_2S\frac{\partial^2V}{\partial S\partial\delta}$ involves both the spot price and convenience yield dimensions and introduces additional numerical complexity compared with one-factor commodity models. Furthermore, the state-dependent diffusion and drift coefficients are what motivate the logarithmic transformation of the spot price introduced in Chapter~\ref{ch:apply}, allowing the PDE to be solved on a more numerically stable computational domain.

For American-style options, the pricing equation is supplemented by the early exercise constraint
\begin{equation}\label{eq:obstacle}
    V(S,\delta,\tau)\geq \psi(S,\delta)=\max(K-S,0)
\end{equation}
which transforms the problem into a free-boundary problem. The numerical scheme must therefore satisfy both the governing PDE and the optimal exercise condition; this formulation is developed in Chapter~\ref{ch:american}.

\subsection{Closed-Form Solution for Futures}\label{sec:closedform}
A significant advantage of the Schwartz two-factor model is the existence of closed-form analytical solutions, which serve as vital benchmarks for numerical schemes. Futures contracts do not possess an early-exercise component, so they satisfy the linear parabolic PDE \eqref{eq:pde_tau} with terminal condition $F(S,\delta,0)=S$. The solution, with $T$ the time to delivery, takes the form
\begin{equation}\label{eq:futures}
F(S,\delta;T)=S\exp\!\big\{-\delta B(T)+A(T)\big\},
\end{equation}
where
\begin{align}
B(T)&=\frac{1-e^{-\kappa T}}{\kappa},\\
A(T)&=\Big(r-\hat\alpha+\frac{\sigma_2^2}{2\kappa^2}-\frac{\rho\sigma_1\sigma_2}{\kappa}\Big)T
+\frac{\sigma_2^2\,(1-e^{-2\kappa T})}{4\kappa^3}
+\Big(\hat\alpha\kappa+\rho\sigma_1\sigma_2-\frac{\sigma_2^2}{\kappa}\Big)\frac{1-e^{-\kappa T}}{\kappa^2}.
\end{align}
Since $S_T=F(S_T,\delta_T;0)$ and $\ln S_T$ is Gaussian under the risk-neutral measure, a European put on the spot price is given by Black's (1976) formula applied to the forward price $F_0=F(S_0,\delta_0;T)$,
\begin{equation}\label{eq:euro}
P_{\mathrm{Eur}}=e^{-rT}\big[K\,\Phi(-d_2)-F_0\,\Phi(-d_1)\big],\qquad d_{1,2}=\frac{\ln(F_0/K)\pm\frac12v^2}{v},
\end{equation}
where $\Phi$ is the standard normal cumulative distribution function and the total variance of $\ln S_T$ is
\begin{equation}\label{eq:var}
v^2=\sigma_1^2T+\frac{\sigma_2^2}{\kappa^2}\Big(T-\frac{2(1-e^{-\kappa T})}{\kappa}+\frac{1-e^{-2\kappa T}}{2\kappa}\Big)-\frac{2\rho\sigma_1\sigma_2}{\kappa}\Big(T-\frac{1-e^{-\kappa T}}{\kappa}\Big).
\end{equation}
The three terms of \eqref{eq:var} are the spot variance, the variance of the integrated convenience yield, and the covariance between the two. The existence of \eqref{eq:euro} allows verification of the ADE solver's accuracy in the European setting before moving to the American free-boundary problem of Chapter~\ref{ch:american}. We use this result to conduct convergence tests on the pricer and ensure that the discretization of the two-dimensional operator correctly captures the interaction between the mean-reverting convenience yield and the underlying price. Equation \eqref{eq:futures} is also used in Chapter~\ref{ch:further}, where the exercise threshold moved to a futures price.

\newpage

\section{The Alternating Direction Explicit Method}\label{ch:ade}

\subsection{Introduction to the ADE Method}
The numerical valuation of American commodity options requires the solution of a two-dimensional, linear parabolic partial differential equation. Implicit finite difference schemes such as Crank--Nicolson (Crank and Nicolson, 1947) and Alternating Direction Implicit (Peaceman and Rachford, 1955) are traditionally favored for their unconditional stability, but they necessitate the solution of large linear systems at each time increment. Thus, when using multi-factor models like the Schwartz (1997) framework, the computational overhead, especially when enforcing the American early-exercise constraint at every grid point, is often significant.\\\\
The ADE method, originally introduced by Saul'yev (1964), provides an explicit finite difference framework for solving parabolic partial differential equations. Unlike conventional explicit schemes, the ADE method uses a discretization in which previously computed values from the current time level are incorporated into each spatial sweep. By updating the solution sequentially across the grid, the method relaxes the stability restrictions of standard explicit schemes while avoiding the large systems of equations required by implicit methods. As a result, ADE combines the implementation simplicity of explicit schemes with many of the stability advantages traditionally associated with the aforementioned implicit methods. The practical use of explicit and implicit finite difference schemes for financial PDEs is surveyed in Duffy (2006, 2022), and these references guide the scheme choices made in this chapter.\\\\
Two forms of the method are considered. The primary scheme in this project is the double-sweep form of Barakat and Clark (1966), in which an upward and a downward sweep are computed at every time step and averaged (Section~\ref{sec:dsweep}). The second is the alternating form of Saul'yev (1964), in which a single sweep is performed per time step and its direction alternates in time (Section~\ref{sec:sauly}). The latter is implemented as a comparison, and the two are evaluated against each other in Chapter~\ref{ch:results}.

\subsection{The Double-Sweep ADE Scheme}\label{sec:dsweep}

The key feature of the double-sweep method is the decomposition of each time step into two complementary directional sweeps. Following the formulation of Barakat and Clark (1966), two intermediate solutions are computed from the same time level, denoted by the upward sweep $U$ and the downward sweep $V$. Unlike conventional explicit finite difference schemes that rely solely on information from the previous time level, each sweep incorporates newly computed values from the current time level as they become available. This sequential ordering preserves the explicit nature of the update while allowing information to propagate through the computational grid within a single time step.\\\\
To illustrate the computational structure, consider the one-dimensional diffusion equation $\frac{\partial u}{\partial t}
=
a\frac{\partial^2u}{\partial x^2}.$  with $\lambda=a\Delta t/h^2$. The upward and downward sweeps are
\begin{align}
\frac{U_j^{n+1}-u_j^n}{\Delta t}&=\frac{a}{h^2}\left(u_{j+1}^{n}-u_j^{n}-U_j^{n+1}+U_{j-1}^{n+1}\right),\label{eq:up1d}\\
\frac{V_j^{n+1}-u_j^n}{\Delta t}&=\frac{a}{h^2}\left(u_{j-1}^{n}-u_j^{n}-V_j^{n+1}+V_{j+1}^{n+1}\right).\label{eq:down1d}
\end{align}
Rearranging gives the computational forms
\begin{align}
(1+\lambda)U_j^{n+1}&=(1-\lambda)u_j^n+\lambda\left(u_{j+1}^{n}+U_{j-1}^{n+1}\right),\\
(1+\lambda)V_j^{n+1}&=(1-\lambda)u_j^n+\lambda\left(u_{j-1}^{n}+V_{j+1}^{n+1}\right).
\end{align}
Although each expression contains a value from the new time level, the update remains explicit because the sweep direction determines which neighbouring value has already been calculated. During the upward sweep, the solution is evaluated sequentially from left to right, making $U_{j-1}^{n+1}$ available when computing $U_j^{n+1}$. The downward sweep follows the same principle in the opposite direction, using previously computed values from the right-hand side of the grid. The final approximation is obtained by averaging the two intermediate solutions,
\begin{equation}\label{eq:avg}
u_j^{n+1}=\frac{1}{2}\left(U_j^{n+1}+V_j^{n+1}\right).
\end{equation}

The asymmetric discretization used in each sweep introduces directional truncation errors. The forward and backward sweeps produce errors of opposite sign, so averaging the two intermediate solutions cancels the leading-order directional error terms and produces a more symmetric approximation. For linear parabolic equations, this averaging procedure results in a discretization that is second-order accurate in space while maintaining favorable numerical properties found in the aforementioned implicit schemes (Duffy, 2022)\\\\

Expanding the bracket in \eqref{eq:up1d} about $(x_j,t_n)$ gives $h^2u_{xx}-\Delta t\,h\,u_{xt}+\dots$, so each sweep approximates
\begin{equation}
u_t=a\,u_{xx}\mp a\frac{\Delta t}{h}\,u_{xt}+\dots,
\end{equation}
with the sign differing between the upward and downward sweeps; this is the directional error that the averaging removes at leading order. Because the remaining error is governed by $\Delta t/h$ rather than $\Delta t/h^2$, the time step must be refined faster than the spatial step for the approximation to remain consistent. This is an important practical property of ADE schemes, and it is visible in the convergence tables of Chapter~\ref{ch:results}.\\\\
Each sweep requires a fixed number of operations per grid point, and the double-sweep scheme performs two sweeps per time step on two arrays. For a grid containing $N$ spatial points and $M$ time steps the computational complexity is therefore $O(MN)$. This linear scaling makes ADE particularly attractive for multi-dimensional problems, where implicit finite difference methods can require significantly more computational effort due to the repeated solving linear systems.

\subsection{The Saul'yev Alternating Scheme}\label{sec:sauly}

Saul'yev's (1964) original scheme uses the same two asymmetric formulae, but applies them \emph{alternately in time} instead of computing both at every step: the upward formula on even time steps and the downward formula on odd time steps,
\begin{equation}
(1+\lambda)u_j^{n+1}=\begin{cases}(1-\lambda)u_j^n+\lambda\left(u_{j+1}^{n}+u_{j-1}^{n+1}\right), & n\ \text{even},\\[2pt](1-\lambda)u_j^n+\lambda\left(u_{j-1}^{n}+u_{j+1}^{n+1}\right), & n\ \text{odd}.\end{cases}
\end{equation}
There is no averaging step, and only one solution array is required, which is overwritten in place so that the freshest value of each neighbor is always used.\\\\
The Saul'yev scheme is likewise unconditionally stable for the diffusion equation (Duffy 2022). The directional error $\mp a(\Delta t/h)u_{xt}$ is not cancelled within a step, but it alternates in sign between consecutive steps and so cancels over a pair of steps, and the same restriction that $\Delta t$ be refined faster than $h$ applies. \\\\
The practical difference between the two variants is therefore one of cost and error cancellation. The Saul'yev scheme performs one sweep and stores one array per time step, roughly half the work and memory of the double-sweep scheme, while the double-sweep scheme cancels the directional error within each step. Chapter~\ref{ch:results} explores whether this trade-off matters for this model.

\subsection{Discretization of the Two-Dimensional Operator}\label{sec:disc2d}

The Schwartz two-factor PDE is a two-dimensional linear parabolic equation containing a cross-derivative term and state-dependent drift coefficients. To apply the ADE method, we write the PDE in the general convection--diffusion--reaction form $V_\tau=\mathcal{L}V$ with
\begin{equation}\label{eq:Lgen}
\mathcal{L}V=a_{11}V_{xx}+2a_{12}V_{xy}+a_{22}V_{yy}+b_1V_x+b_2V_y-rV,
\end{equation}
where the specific coefficients for the Schwartz model are given in Chapter~\ref{ch:apply}. On a uniform grid $(x_i,y_j)$ with spacings $h_x,h_y$ and time step $\Delta\tau$, define
\begin{equation}
c_x=\frac{a_{11}}{h_x^2},\qquad c_y=\frac{a_{22}}{h_y^2},\qquad c_{xy}=\frac{a_{12}}{2h_xh_y},\qquad \beta=c_x+c_y,
\end{equation}
and the four-point difference approximating the cross derivative, evaluated at the old time level
\begin{equation}
\delta_{xy}u^n_{i,j}=u^n_{i+1,j+1}-u^n_{i+1,j-1}-u^n_{i-1,j+1}+u^n_{i-1,j-1}.
\end{equation}
The second derivatives in each direction use the sweep construction of \eqref{eq:up1d} and \eqref{eq:down1d}, the first derivatives use the mixed-level central difference $\frac{1}{2h}(u^{n}_{i+1}-U^{n+1}_{i-1})$ in the upward sweep, and the reaction term is treated explicitly at the old time level. For the double-sweep scheme, the upward sweep (loops over $i$ and $j$ increasing) is
\begin{align}
(1+\Delta\tau\,\beta)\,U_{i,j}^{n+1}={}&\big(1-\Delta\tau(\beta+r)\big)\,u_{i,j}^n+\Delta\tau\Big[c_x\big(u_{i+1,j}^{n}+U_{i-1,j}^{n+1}\big)+c_y\big(u_{i,j+1}^{n}+U_{i,j-1}^{n+1}\big)\notag\\
&+\frac{b_1}{2h_x}\big(u_{i+1,j}^{n}-U_{i-1,j}^{n+1}\big)+\frac{b_2}{2h_y}\big(u_{i,j+1}^{n}-U_{i,j-1}^{n+1}\big)+c_{xy}\,\delta_{xy}u^n_{i,j}\Big],\label{eq:up2d}
\end{align}
and the downward sweep exchanges the roles of the two time levels,
\begin{align}
(1+\Delta\tau\,\beta)\,V_{i,j}^{n+1}={}&\big(1-\Delta\tau(\beta+r)\big)\,u_{i,j}^n+\Delta\tau\Big[c_x\big(u_{i-1,j}^{n}+V_{i+1,j}^{n+1}\big)+c_y\big(u_{i,j-1}^{n}+V_{i,j+1}^{n+1}\big)\notag\\
&+\frac{b_1}{2h_x}\big(V_{i+1,j}^{n+1}-u_{i-1,j}^{n}\big)+\frac{b_2}{2h_y}\big(V_{i,j+1}^{n+1}-u_{i,j-1}^{n}\big)+c_{xy}\,\delta_{xy}u^n_{i,j}\Big].\label{eq:down2d}
\end{align}
The solution is then $u^{n+1}_{i,j}=\tfrac12(U^{n+1}_{i,j}+V^{n+1}_{i,j})$ as in \eqref{eq:avg}. The Saul'yev variant uses exactly the same two update formulae, but with a single array $u^{n+1}$ in place of $U^{n+1}$ and $V^{n+1}$: \eqref{eq:up2d} is applied on even steps, \eqref{eq:down2d} on odd steps, and no averaging is performed. In both cases the cross-derivative term and the middle value use the previous time level, and the reaction term is explicit, so only the neighboring values in the sweep direction are taken from the new level.\\\\
The stability analysis above applies to the diffusion terms. The convection, reaction and cross-derivative terms are treated at lower order and are not covered by that argument, so their effect on stability is verified numerically.

\newpage

\section{Applying the Numerical Scheme}\label{ch:apply}

\subsection{Transforming the Computational Domain}

The implementation of the ADE method requires transforming the continuous Schwartz two-factor pricing equation into a finite computational framework. Since both state variables evolve over theoretically unbounded domains, direct discretization in the original variables would require an unnecessarily large computational region. Therefore, coordinate transformations are introduced to obtain a bounded numerical domain while preserving the structure of the underlying stochastic process.\\\\
Extending Section~\ref{sec:pde}, the spot price is transformed using the logarithmic variable $x=\ln S$. Applying It\^o's lemma to \eqref{eq:S_rn} gives
\begin{equation}
 dx=\left(r-\delta-\frac{1}{2}\sigma_1^2\right)dt+\sigma_1dz_1
\end{equation}
which removes the state dependence from the diffusion coefficient. This transformation improves numerical stability by allowing the computational grid to provide a more uniform resolution across different orders of magnitude of the spot price. The convenience yield is kept as the second coordinate, $y=\delta$, and each variable is restricted to a finite interval,
\[
(x,y)\in[x_{\min},x_{\max}]\times[y_{\min},y_{\max}],
\]
producing a rectangular computational domain on which the ADE sweeps can be applied using a uniform spatial mesh.The domain is centred on the initial state $(S_0,\delta_0)$ of each scenario, so that the initial point is the centre node of the mesh,
\begin{equation}
x\in[\ln S_0-1.2,\ \ln S_0+1.2],\qquad y\in[\delta_0-0.4,\ \delta_0+0.4],
\end{equation}
with $N_X$ and $N_Y$ intervals ($N_X=N_Y$ in all experiments), giving spacings $h_x=2.4/N_X$ and $h_y=0.8/N_Y$. In the spot variable the first interval runs from $S_0e^{-1.2}$ to $S_0e^{1.2}$, about $0.30S_0$ to $3.3S_0$, which is roughly $10.7$ standard deviations of $\ln S_T$ on each side, since $\sigma_1\sqrt{T}\approx0.112$ for the parameters of Section~\ref{sec:params}. In the yield variable the stationary standard deviation of the Ornstein--Uhlenbeck process is $\sigma_2/\sqrt{2\kappa}\approx0.067$, so the half-width of $0.4$ is about six stationary standard deviations. The interval is centred on $\delta_0$ rather than $\hat\alpha$, so that the initial yield lies at the centre node; for example it is $[-0.3,0.5]$ for $\delta_0=0.1$ and $[-0.1,0.7]$ for $\delta_0=0.3$. These widths make the probability of the process reaching an artificial edge during the life of the option negligible.

\subsection{Transformation of the Schwartz PDE}

Under the coordinate transformations introduced above, the Schwartz PDE takes the general form \eqref{eq:Lgen} in $\mathbf{z}=(x,y)^\top$ with
\begin{equation}\label{eq:transformedPDE}
V_\tau=\tfrac12\sigma_1^2V_{xx}+\rho\sigma_1\sigma_2V_{xy}+\tfrac12\sigma_2^2V_{yy}+\left(r-y-\tfrac12\sigma_1^2\right)V_x+\kappa(\hat\alpha-y)V_y-rV.
\end{equation}
The transformed diffusion tensor and drift coefficients are therefore
\begin{equation}
\hat a=\begin{pmatrix}\tfrac12\sigma_1^2&\tfrac12\rho\sigma_1\sigma_2\\[2pt]\tfrac12\rho\sigma_1\sigma_2&\tfrac12\sigma_2^2\end{pmatrix},\qquad
\hat b=\begin{pmatrix}r-y-\tfrac12\sigma_1^2\\ \kappa(\hat\alpha-y)\end{pmatrix},
\end{equation}
that is, $a_{11}=\frac12\sigma_1^2$, $a_{12}=\frac12\rho\sigma_1\sigma_2$, $a_{22}=\frac12\sigma_2^2$, $b_1=r-y-\frac12\sigma_1^2$ and $b_2=\kappa(\hat\alpha-y)$ in \eqref{eq:Lgen}. All diffusion coefficients are now constants, and only the drifts depend on the state, through $y$.

The coordinate transformation retains the parabolic structure of the original equation while producing coefficients suitable for numerical discretization on a finite grid. The transformed operator consists of four components: the diffusion terms associated with the two stochastic factors, the cross-derivative term generated by the correlation between the processes, the convection terms arising from the drift components, and the reaction term associated with discounting.

\subsection{Boundary Treatment via Fichera Theory}\label{sec:fichera}

The transformation of the Schwartz PDE produces a finite computational domain, however boundaries must be introduced to approximate the behavior of the solution outside this region. The treatment of these boundaries is important particularly in regions where the diffusion tensor becomes degenerate. Rather than imposing boundary conditions uniformly across the computational domain, the appropriate treatment is determined by the direction of information flow through the boundary.\\\\
At boundaries where the diffusion tensor becomes degenerate, this behavior is analyzed using Fichera theory (Fichera, 1960). The Fichera function is defined as
\begin{equation}
\mathcal{F}(\mathbf{x})=\Big(b_i-\sum_j\frac{\partial a_{ij}}{\partial x_j}\Big)n_i,
\end{equation}
where $a_{ij}$ denotes the diffusion tensor, $b_i$ the drift vector, and $n_i$ the inward unit normal vector. The sign of the Fichera function determines whether additional boundary information must be supplied or whether the PDE itself determines the solution behavior at the boundary. Imposing artificial conditions at boundaries where information naturally leaves the domain may over-constrain the PDE and introduce numerical inaccuracies.

This analysis is applied in the original variables $(S,\delta)$, where the degeneracy is visible. As $S\to0$, the diffusion coefficients $a_{11}=\frac12\sigma_1^2S^2$ and $a_{12}=\frac12\rho\sigma_1\sigma_2S$ vanish, and the Fichera function for the inward normal $n=(1,0)$ is
\begin{equation}
\mathcal{F}\big|_{S=0}=(r-\delta)S-\sigma_1^2S-\frac{\partial a_{12}}{\partial\delta}\Big|_{S=0}=0 .
\end{equation}
The lower spot boundary is therefore characteristic: the PDE reduces there to the first-order equation
\begin{equation}
V_\tau=\kappa(\hat\alpha-\delta)V_\delta+\tfrac12\sigma_2^2V_{\delta\delta}-rV,
\end{equation}
whose solution is known from the financial behavior of the option. For a European put, $V\to Ke^{-r\tau}$, which satisfies this equation. For an American put the holder exercises immediately at $S=0$, so $V\to K$, while for a call the value approaches zero. A Dirichlet condition is therefore imposed at the truncated boundary $S_{\min}$, selected close to zero such that the limiting approximation remains accurate. In the logarithmic variable this edge corresponds to $x\to-\infty$ and is represented by $x_{\min}$.

The opposite extreme of the spot-price domain is treated through the asymptotic behavior of the option. As $S\to\infty$ the value of a put option approaches zero, giving $V(x_{\max},y,\tau)=0$. These conditions ensure that the finite computational domain remains consistent with the theoretical behavior of the option value outside the grid.\\\\
Unlike the spot-price process, the convenience yield follows a mean-reverting Ornstein--Uhlenbeck process and therefore possesses an unbounded theoretical domain with no natural absorbing boundary. The variable is restricted to a sufficiently large finite interval, and at its edges the solution is extrapolated linearly,
\begin{equation}
V_{yy}=0\quad\Longrightarrow\quad u_{i,0}=2u_{i,1}-u_{i,2},\qquad u_{i,N_y}=2u_{i,N_y-1}-u_{i,N_y-2},
\end{equation}
a standard choice when the edge lies far from the region of interest. This treatment preserves the stochastic characteristics of the Schwartz model while allowing the ADE method to operate efficiently on a finite computational mesh. The boundary values are set at the new time level before the sweeps begin and refreshed after them, so that every sweep uses current boundary data.

\newpage

\section{American Option Application}\label{ch:american}

\subsection{The American Option Valuation Problem}

The introduction of early exercise fundamentally changes the numerical structure of the option pricing problem. Unlike European options, where the holder can exercise only at maturity, American options allow exercise at any point before expiration. Consequently, the option value is determined by a trade-off between immediate exercise and continuation of the contract.

Within the Schwartz two-factor framework, this decision depends on both the current spot price and the stochastic convenience yield. The optimal exercise strategy is therefore not known prior to solving the pricing problem and must be determined simultaneously with the option value. This introduces a free boundary separating the exercise region
\begin{equation}
\mathcal{E}=\{(S,\delta,\tau):V=\psi\},
\end{equation}
where immediate exercise is optimal, from the continuation region $\mathcal{C}=\{V>\psi\}$, where retaining the option provides greater value. For a put, the exercise surface can be written as $S^*(\delta,\tau)=\sup\{S:V(S,\delta,\tau)=\psi(S,\delta)\}$, a nontrivial surface in the state space $(S,\delta,\tau)$ rather than a single threshold. The numerical challenge is therefore not only to solve the governing PDE but also to accurately determine the evolving location of this surface. This feature is particularly important for commodity options, where stochastic convenience yields can significantly influence incentives for early exercise through their effect on the benefits of holding physical inventory (Gibson and Schwartz, 1990; Hilliard and Reis, 1998).

\subsection{Variational Inequality Formulation}

The American option problem can be expressed as a variational inequality that combines the original pricing equation with the constraint that the option value cannot fall below its immediate exercise value. Let $\psi(S,\delta)$ denote the payoff function. The option value satisfies
\begin{equation}\label{eq:lcp}
V_\tau-\mathcal{L}V\geq 0,\qquad
V-\psi\geq 0,\qquad
\left(V_\tau-\mathcal{L}V\right)(V-\psi)=0,\qquad V(\cdot,0)=\psi.
\end{equation}
The first condition represents the pricing equation in the continuation region, while the second condition enforces the dominance of the option value over the immediate exercise payoff. The final complementarity condition ensures that only one of these conditions is active at any point in the domain.\\\\
When the option lies in the continuation region, $V>\psi$, and the governing Schwartz PDE is satisfied as an equality. Conversely, when immediate exercise is optimal, $V=\psi$, and the value function is constrained by the payoff surface. Therefore, the American option problem can be interpreted as a constrained PDE problem, where the numerical solution must remain above a threshold defined by the payoff function.

\subsection{Discrete Enforcement of the Exercise Constraint}

The ADE formulation developed in the previous chapters provides an approximation of the continuation value at each time step. To incorporate the American exercise feature, the unconstrained ADE solution is projected onto the admissible region after the sweeps (and after the averaging step in the double-sweep scheme). At a given grid point $(x_i,y_j)$, the numerical solution is modified according to
\begin{equation}\label{eq:projection}
u_{i,j}^{n+1}=\max\left(u_{i,j}^{n+1,*},\ \psi(x_i,y_j,\tau_{n+1})\right),
\end{equation}
where $u_{i,j}^{n+1,*}$ represents the value obtained from the ADE update (Section~\ref{sec:disc2d}) before applying the exercise constraint. This projection step enforces the inequality condition $u_{i,j}^{n+1}\geq\psi(x_i,y_j)$ at every point in the computational domain. If the unconstrained solution falls below the payoff function, the grid point is classified as belonging to the exercise region and is reset to the immediate exercise value. Otherwise, the point remains in the continuation region and retains the PDE solution.

\subsection{Advantages of ADE for American Option Pricing}

Traditional finite difference approaches for American options commonly formulate the discrete problem as a Linear Complementarity Problem (LCP). Implicit finite difference schemes require the solution of a coupled linear system at each time step while simultaneously enforcing the inequality constraints. This often requires iterative procedures such as Projected Successive Over-Relaxation (PSOR), increasing computational complexity as the dimensionality of the problem grows (Duffy, 2006).\\\\
With ADE the underlying finite difference update remains explicit, and the early-exercise constraint is enforced locally through \eqref{eq:projection}. For a computational grid containing $N_x$ and $N_y$ points in the two state variables, each time step requires operations proportional to the number of grid points,
\begin{equation}
\text{cost per step}=O(N_xN_y),\qquad \text{total cost}=O(N_xN_yN_\tau).
\end{equation}
Therefore, the addition of the American exercise constraint does not alter the fundamental computational complexity of the ADE solver. This efficiency is particularly valuable for two-factor commodity models, where implicit approaches require significantly greater computational effort due to the increased dimensionality of the resulting linear systems.

\subsection{Numerical Treatment of the Exercise Boundary}

The accuracy of an American option approximation depends strongly on the ability of the numerical scheme to resolve the transition between the exercise and continuation regions. Under appropriate regularity conditions, the optimal boundary satisfies the smooth pasting condition, meaning that both the option value and its first derivative match the payoff function at the exercise boundary,
\begin{equation}
V\big|_{S^*}=\psi\big|_{S^*},\qquad \nabla V\big|_{S^*}=\nabla\psi\big|_{S^*}.
\end{equation}
In the Schwartz two-factor model, the exercise boundary is influenced by both the spot price and convenience yield, producing a nontrivial surface rather than a one-dimensional threshold. Accurate resolution of this surface therefore requires a numerical method capable of maintaining stability while capturing sharp changes in the solution gradient.\\\\
The alternating sweep structure of the ADE method reduces directional bias. In the double-sweep scheme, information propagated from opposite spatial directions is combined within each step, and in the Saul'yev variant the bias of one step is offset by the next. Either way, the resulting solution provides a more symmetric approximation near the free boundary than a scheme with a fixed sweep direction. The projection step ensures that the computed value remains consistent with the American exercise constraint. Together, these features allow the ADE framework to efficiently approximate American commodity option values under stochastic convenience yield dynamics.

\newpage

\section{Computation and Validation Methods}\label{ch:validation}

\subsection{Implementation of the Numerical Solver}

The numerical framework developed in this study implements the transformed Schwartz two-factor PDE \eqref{eq:transformedPDE} using the ADE schemes described in the previous chapters, with the double-sweep form as the primary scheme and the Saul'yev alternating form implemented as an alternative sweep routine for comparison. The solver is developed in C++ to provide computational efficiency while maintaining a modular structure between the financial model, numerical discretization, and computational grid. It uses the object-oriented approach described by Duffy (2004), including many of its base files in the code. The implementation separates the specification of the PDE coefficients, boundary conditions, payoff functions, and finite difference operations from the underlying solver algorithm, and it utilizes matrix and vector structures from Duffy (2004) designed to correspond with the mathematical notation of the finite difference formulation. This reduces indexing errors and allows the numerical approximation of the Schwartz operator to be directly represented within the computational framework.\\\\
Each time step $n=0,\dots,N_\tau-1$ of the double-sweep solver performs the following operations:
\begin{enumerate}
\item set the boundary values at $\tau_{n+1}$ (Section~\ref{sec:fichera});
\item compute the upward sweep $U^{n+1}$ with \eqref{eq:up2d} and, separately, the downward sweep $V^{n+1}$ with \eqref{eq:down2d}, both from $u^n$;
\item average, $u^{n+1}=\tfrac12(U^{n+1}+V^{n+1})$, and refresh the boundary values;
\item for American options, project onto the payoff using \eqref{eq:projection}.
\end{enumerate}
The Saul'yev variant replaces steps 2 and 3 by a single in-place sweep, upward with \eqref{eq:up2d} if $n$ is even and downward with \eqref{eq:down2d} if $n$ is odd, and requires an even number of time steps $N_\tau$ so that every upward sweep is paired with a downward sweep. The same numerical engine evaluates both European and American options, with the only modification being step 4. The sweep routines of both schemes are given in the accompanying source file.

\subsection{European Option Benchmarking}

Before evaluating the American option problem, the accuracy of the underlying ADE discretization is verified as a European option. Removing the early-exercise constraint isolates errors arising from the finite difference approximation, coordinate transformations, and treatment of the cross-derivative term. The solver's finest-grid result is compared with two independent benchmarks.\\\\
The first is the closed-form price \eqref{eq:euro}. The second is a Monte Carlo estimate over $M$ simulated paths of the joint process \eqref{eq:S_rn}, \eqref{eq:d_rn2},
\begin{equation}
\hat P=\frac{e^{-rT}}{M}\sum_{m=1}^{M}\max\big(K-S_T^{(m)},0\big),\qquad \mathrm{SE}=\frac{e^{-rT}s_M}{\sqrt M},
\end{equation}
where $s_M$ is the sample standard deviation of the discounted payoffs. Consistency between these three solutions provides verification that the transformed PDE coefficients, boundary treatment, and discretization of the two-factor diffusion process have been implemented correctly before introducing the additional complexity of early exercise.

\subsection{American Option Benchmarking}

The valuation of American options provides a more challenging test of the numerical framework because the exercise boundary is unknown and must be determined as part of the solution. Since no closed-form solution generally exists for American options under the Schwartz two-factor model, an independent numerical benchmark is required.\\\\
The Least Squares Monte Carlo (LSMC) method introduced by Longstaff and Schwartz (2001) is used as a comparison method. Unlike the ADE approach, which solves the problem through spatial discretization of the pricing PDE, LSMC determines the optimal exercise strategy by simulating paths of the state variables and working backward through the exercise dates. At each date $t_k$, over the paths that are in the money, the continuation value is approximated by regression,
\begin{equation}
C(S,\delta)\approx\sum_{l=1}^{L}\beta_l\,\phi_l(S,\delta),
\end{equation}
where the coefficients $\beta_l$ are obtained by ordinary least squares on the discounted cash flows of the later dates, and the path is exercised if the immediate payoff exceeds the fitted continuation value. The basis functions $\phi_l$ are weighted Laguerre polynomials; for the two-factor Schwartz model, this basis includes terms for the spot price, the convenience yield, and their cross-products.\\\\
The ADE and LSMC approaches therefore provide complementary valuation methods with different sources of numerical error: discretization error for ADE, and regression and sampling error for LSMC. Agreement between the two methods shows that the implementation of the early-exercise constraint and the approximation of the free boundary are accurate. Additionally, the comparison highlights the computational efficiency gained through the PDE-based ADE approach relative to simulation-based methods.

\newpage

\section{Numerical Results}\label{ch:results}

\subsection{Parameters and Inputs}\label{sec:params}

The numerical experiments are conducted using the two-factor Schwartz model parameters reported in Schwartz (1997). These parameters were estimated using Kalman filtering techniques (briefly mentioned in Chapter~\ref{ch:model}) applied to historical commodity price data and provide a representation of WTI crude oil dynamics. The parameter set used throughout the experiments is
\[\kappa = 1.49, \,\hat{\alpha} = 0.0161,\,\sigma_1 = 0.158,\, \sigma_2=0.115,\,\rho=0.189,\,r=0.05\]
A strike price of $K=100$ and time to maturity of $T=0.5$ is selected to represent a realistic current-day price environment for crude oil.\\\\
In ~\ref{sec:season}, we will explore the addition of seasonality, which makes the parameters changing with time, rather than as point estimates. 

\subsection{European Option Testing}

We use an initial spot price of $S_0=90$ and corresponding initial convenience yield $\delta_0 = 0.1$. This produces an initially in-the-money put option and provides a meaningful test of the early-exercise constraint.\\\\
Before evaluating the American option implementation, the ADE solver is first validated in the European setting by removing the early-exercise constraint and comparing the numerical approximation against the analytical Schwartz solution \eqref{eq:euro}. The analytical solution provides a benchmark price of $11.601861$, while an independent Monte Carlo simulation produces an estimate of $11.602859$ with a standard error of $0.0164725$. The ADE approximation produces a price of $11.601279$, differing from the analytical solution by approximately $5.82\times10^{-4}$. The consistency between these three approaches confirms that the transformed Schwartz PDE, coordinate transformations, and finite difference discretization have been implemented correctly before introducing the additional complexity of the American exercise constraint.

\subsection{American Option and Convergence Testing}
\begin{table}
\centering
\caption{Double-sweep ADE American Put Price $S_0 =90$, $\delta_0 = 0.1$}
\label{tab:table1}
\begin{tabular}{c|cccc}
\hline
$N_Y \backslash N_T$ & 100 & 300 & 600 & 1200\\
\hline
50  & 11.8684 & 11.8711 & 11.8715 & 11.8716\\
100 & 11.8821 & 11.8898 & 11.8907 & 11.8909\\
200 & 11.8598 & 11.8880 & 11.8908 & 11.8916\\
400 & 11.7705 & 11.8788 & 11.8894 & 11.8921\\
\hline
\end{tabular}
\end{table}
The convergence results of the double-sweep scheme in Table~\ref{tab:table1} demonstrate that the ADE solution stabilizes as both spatial and temporal resolutions are increased, with the finest grid producing an American option value of approximately $11.892050$. Temporal refinement produces rapid convergence, while additional spatial refinement improves the approximation of the early-exercise boundary once the time step is small enough. For the coarsest time grid ($N_T=100$), however, refining the spatial grid beyond $N_Y=100$ reduces the price from $11.8821$ to $11.7705$. This is the behavior predicted in Section~\ref{sec:dsweep}: the ADE error depends on $\Delta\tau/h$ rather than $\Delta\tau/h^2$, so decreasing $h$ without decreasing $\Delta\tau$ increases the error. The practical consequence is that time steps should be refined at least as fast as spatial steps when using ADE.
\\\\ As mentioned in Chapter~\ref{ch:validation}, since no analytical solution exists for American options under the Schwartz two-factor model, we benchmark with 250,000 Monte Carlo paths under LSMC, which estimates $11.888911$ with a standard error of $0.014396$. The difference between the two approaches is approximately $0.0031$, which remains well within the uncertainty of the LSMC estimate. This consistency demonstrates the numerical robustness of the proposed ADE framework.

\subsection{Scenario Analysis of American Commodity Options}

To evaluate the performance of the ADE solver under different commodity market conditions, three initial state scenarios are considered. The first scenario represents the base case with an in-the-money put option, where the spot price is below the strike and the initial convenience yield is consistent with the original calibration. The second scenario examines an at-the-money option by increasing the spot price to the strike level while keeping the convenience yield unchanged. This case evaluates the behavior of the numerical scheme near the transition between the continuation and exercise regions. The final scenario investigates the impact of an elevated convenience yield, representing a market environment with stronger benefits from holding physical inventory. In all cases, the ADE results are benchmarked against the LSMC estimates. The resulting option values are summarized in Table~\ref{tab:Table2}.

\begin{table}[h!]
\centering
\caption{American put option values under different initial commodity states.}
\label{tab:Table2}
\begin{tabular}{cccccc}
\hline
$S_0$ & $\delta_0$ & Scenario & ADE Price & LSMC Price & Absolute Difference \\
\hline
90  & 0.1 & ITM & 11.892050 & 11.888911 & 0.003139 \\
100 & 0.1 & ATM & 5.020935  & 5.024249  & 0.003314 \\
90  & 0.3 & ITM, High Yield & 16.952677 & 16.989201 & 0.036524 \\
\hline
\end{tabular}
\end{table}

The results demonstrate strong consistency between the ADE and LSMC approaches across all three market scenarios. For the base in-the-money case, the ADE estimate differs from the LSMC benchmark by only $0.003139$, a relative error of approximately $0.03\%$. The at-the-money scenario produces a similarly small deviation of $0.003314$, indicating that the ADE scheme accurately captures the behavior of the solution near the exercise boundary, where numerical methods are typically most sensitive.\\\\
The final scenario highlights the influence of the convenience yield on American commodity option values. Increasing the initial convenience yield from $0.1$ to $0.3$ increases the option value from approximately $11.89$ to $16.95$. This is consistent with the futures formula \eqref{eq:futures}: the risk-neutral expected spot price at maturity is the forward price,
\begin{equation}
E^{\mathbb{Q}}[S_T]=F(S_0,\delta_0;T)=S_0\,e^{-\delta_0B(T)+A(T)},
\end{equation}
which decreases in $\delta_0$ (for $T=0.5$, raising $\delta_0$ from $0.1$ to $0.3$ lowers the forward by a factor $e^{-0.2B(T)}\approx0.93$), and a lower expected terminal spot raises the value of a put. This behavior is consistent with commodity option theory such as in Pirrong (2012), as a higher convenience yield reflects greater benefits from holding physical inventory relative to holding a claim on it. The difference between the ADE and LSMC estimates increases slightly in this scenario to $0.036524$ (about $0.2\%$ of the option value), although the discrepancy remains small relative to the option value. This suggests that the ADE method remains robust when the exercise boundary becomes more sensitive to changes in the underlying commodity state.\\\\
Overall, the scenario analysis confirms that the ADE framework accurately prices American options across different regions of the state space while maintaining consistency with the independent LSMC benchmark.

\subsection{Comparison of the Double-Sweep and Saul'yev Schemes}\label{sec:compare}
 
The Saul'yev alternating variant (Section~\ref{sec:sauly}) is run for the same $(N_Y,N_T)$ combinations as Table~\ref{tab:table1}, with $N_X=N_Y$, and the two schemes are compared on their outputs  Table~\ref{tab:compare} reports the American put price at $S_0=90$, $\delta_0=0.1$ and the difference between the schemes. Computational cost is treated separately in Section~\ref{sec:runtime}.
 
\begin{table}[h!]
\centering
\caption{Double-sweep versus Saul'yev ADE, American put, $S_0=90$, $\delta_0=0.1$.}
\label{tab:compare}
\begin{tabular}{cc|ccc}
\hline
$N_Y$ & $N_T$ & Double sweep & Saul'yev & Difference (S $-$ DS) \\
\hline
100 & 300 & 11.889786 & 11.889802 & $1.6\times10^{-5}$ \\
100 & 1200 & 11.890940 & 11.890941 & $1.2\times10^{-6}$ \\
200 & 300 & 11.888018 & 11.888039 & $2.1\times10^{-5}$ \\
200 & 1200 & 11.891584 & 11.891587 & $2.3\times10^{-6}$ \\
\hline
\end{tabular}
\end{table}
 
The two schemes agree to within $2.1\times10^{-5}$ on every grid, which is less than $0.2\%$ of the LSMC standard error of $0.0144$ reported above. The difference is largest at the coarse time grid ($N_T=300$) and shrinks by roughly an order of magnitude at $N_T=1200$, so both schemes converge to the same limit and the choice between them does not change the price in any material way. Estimating the order of convergence in $\Delta\tau$ from three successive refinements, $p=\log_2\!\big[(V_{600}-V_{300})/(V_{1200}-V_{600})\big]$, gives $p=1.65$ (double sweep) and $1.64$ (Saul'yev) at $N_Y=100$, and $1.88$ for both at $N_Y=200$. This two schemes also look to converge at nearly the same rate, which further strengthens the validity of the results.
  
\subsection{Computational Efficiency and Runtime}\label{sec:runtime}
 
A principal motivation for ADE is cost, so we compare the runtime time of the double-sweep scheme, the Saul'yev scheme and LSMC under the same conditions in C++.
The PDE timings are the best of three runs of the full solve, including mesh construction and the American projection, and the LSMC timing is a single run with $50$ exercise dates, a weighted Laguerre regression basis (Chapter~\ref{ch:validation}), and $M$ simulated paths including path generation. The accuracy of the ADE prices is measured against the reference value $V_{\mathrm{ref}}=11.892754$, the double-sweep price on a finer grid than run in prior runs ($N_X=N_Y=400$, $N_T=2400$); this measures convergence of the scheme rather than modelling error, which is checked separately against the closed form and LSMC. The accuracy of the LSMC prices is reported as the standard error.\\\\
For $s$ sweeps per time step, the ADE cost is
\begin{equation}
T_{\mathrm{ADE}}\approx c_{\mathrm{sw}}\,s\,N_xN_yN_\tau,\qquad s=2\ \text{(double sweep)},\quad s=1\ \text{(Saul'yev)},
\end{equation}
where $c_{\mathrm{sw}}$ is the cost of one grid-point update, so the double-sweep scheme should take about twice as long as the Saul'yev scheme on the same grid. For LSMC, the cost is $O(MN\,n_b^2)$ for the regressions plus $O(MN)$ for path generation\\\\ 
\begin{table}[h!]
\centering
\caption{Runtime and accuracy of the double-sweep ADE, Saul'yev ADE and LSMC (American put, $S_0=90$, $\delta_0=0.1$). For ADE the error is $|V-V_{\mathrm{ref}}|$; for LSMC the entry is the standard error. Best of 3 runs for ADE, single run for LSMC.}
\label{tab:runtime}
\begin{tabular}{llcc}
\hline
Method & Configuration & $|$Error$|$ / SE & Time (s) \\
\hline
Double sweep & $N_Y=50,\ N_T=300$ & $2.2\times10^{-2}$ & 0.093 \\
Saul'yev & $N_Y=50,\ N_T=300$ & $2.2\times10^{-2}$ & 0.050 \\
Double sweep & $N_Y=100,\ N_T=600$ & $2.1\times10^{-3}$ & 0.749 \\
Saul'yev & $N_Y=100,\ N_T=600$ & $2.1\times10^{-3}$ & 0.410 \\
Double sweep & $N_Y=200,\ N_T=600$ & $1.9\times10^{-3}$ & 2.986 \\
Saul'yev & $N_Y=200,\ N_T=600$ & $1.9\times10^{-3}$ & 1.655 \\
Double sweep & $N_Y=200,\ N_T=1200$ & $1.2\times10^{-3}$ & 6.016 \\
Saul'yev & $N_Y=200,\ N_T=1200$ & $1.2\times10^{-3}$ & 3.267 \\
\hline
LSMC & $M=10000$ & $\pm0.0708$ & 0.088 \\
LSMC & $M=50000$ & $\pm0.0327$ & 0.632 \\
LSMC & $M=250000$ & $\pm0.0144$ & 3.365 \\
\hline
\end{tabular}
\end{table}
 
A couple of interesting conclusions can be drawn rom Table~\ref{tab:runtime}. First, it shows the Saul'yev scheme is between $1.80$ and $1.89$ times faster than the double-sweep scheme on every grid tested, in line with the cost model above. Table~\ref{tab:compare} shows that it returns the same price to within $2.1\times10^{-5}$. Since the schemes converge at the same rate, the single-sweep variant gives the same accuracy for roughly half the cost, and the double-sweep scheme's extra sweep does not buy a measurable accuracy gain in this problem.\\\\
Importnatly, it is evident that ADE is far cheaper than LSMC at equal accuracy, and the advantage grows as the accuracy requirement tightens. The LSMC run with $M=250{,}000$ paths takes $3.4$\,s and has a standard error of $0.0144$. The double-sweep run at $N_Y=200$, $N_T=600$ takes a comparable $3.0$\,s but has an error of $1.9\times10^{-3}$, about $7.6$ times smaller, and the Saul'yev scheme achieves the same accuracy in $1.7$\,s. A coarser grid ($N_Y=100$, $N_T=600$) already reaches an error of $2.1\times10^{-3}$ in $0.75$\,s (double sweep) or $0.41$\,s (Saul'yev). Matching that accuracy with LSMC would require about $(0.0144/0.0021)^2\approx47$ times as many paths, that is about $1.2\times10^{7}$ paths and, since LSMC time scales linearly in $M$, an extrapolated run time of about $160$\,s. On this basis ADE is roughly $210$ times faster than LSMC for the double-sweep scheme and roughly $390$ times faster for the Saul'yev scheme at an accuracy of about $2\times10^{-3}$. This exploration shows that ADE consistently reaches a given accuracy with significantly faster runtimes than LSMC.

\newpage

\section{Practical Extensions and Further Study}\label{ch:further}

Following feedback from a commodities trader specializing in agricultural markets, four extensions of the framework were tested as proof of concept. The feedback recognized this project as a numerical extension of Schwartz (1997) rather than a new economic model, and suggested extensions that would bring the pricer closer to desk use. All of them are numerical modifications of the same solver, and the results reported here are preliminary.

\subsection{Exercise Obstacle on a Futures Contract}

Listed options on agricultural commodities are written on a specific futures delivery month rather than directly on spot, and spot is itself often not a single clean price because of location, quality and basis effects. We therefore replace the spot obstacle by one on a futures price. Let $T_F\ge T$ be the delivery date of the underlying contract and $\theta=T_F-T$ its residual life at option expiry. At option time-to-maturity $\tau$, the futures contract has time $\tau+\theta$ to delivery, so by \eqref{eq:futures} the obstacle becomes
\begin{equation}
\psi_F(x,\delta,\tau)=\max\Big\{K-F\big(e^x,\delta;\tau+\theta\big),\,0\Big\}=\max\Big\{K-e^{\,x-\delta B(\tau+\theta)+A(\tau+\theta)},\,0\Big\},
\end{equation}
with initial condition $V(\cdot,0)=\psi_F(\cdot,0)$. State variables remain $S$ and $\delta$. Two features distinguish this from the spot problem. First, the obstacle depends on time through $\tau$. Second, the exercise region is a nonlinear surface in $(S,\delta)$, because the futures price depends on the convenience yield through the factor $e^{-\delta B(\tau+\theta)}$, so the exercise decision depends on the full state rather than a spot threshold. Since $\psi_F$ is evaluated pointwise, only the projection step \eqref{eq:projection} changes; the ADE sweeps are unaffected. The contractual exercise and settlement convention of an actual contract would need to be matched in a market application.

\subsection{Seasonal Parameters}\label{sec:season}

For storable agricultural commodities, planting, harvest and storage cycles suggest seasonality in the risk-adjusted convenience yield level and in the volatilities. In the solver, the constant parameters are replaced by periodic functions of the time to maturity $\tau$,
\begin{align}
\hat\alpha(\tau)&=0.0161+0.005\,\sin(2\pi\tau),\notag\\
\sigma_1(\tau)&=0.158+0.03\,\cos(2\pi\tau),\label{eq:seasonal}\\
\sigma_2(\tau)&=0.115+0.03\,\lvert\sin(\pi\tau)\rvert,\notag
\end{align}
where the sine and cosine are chosen to represent the inverse relationship between the planting and harvest cycles, and the constant terms are the Schwartz (1997) estimates. At each time step $n$ the coefficients are evaluated at $\tau_n=n\Delta\tau$ and held fixed during that step. All three functions remain positive for the amplitudes used. The amplitudes are illustrative rather than calibrated.\\\\
Because the option has $T=0.5$, only half a period of each function is traversed, so their time averages differ from the constants. Over $\tau\in[0,0.5]$ the average of $\hat\alpha$ is raised by about $0.003$ and that of $\sigma_2$ by about $0.019$, while the average of $\sigma_1$ is unchanged at $0.158$ even though $\sigma_1$ varies between $0.128$ and $0.188$. Including seasonality increased the computed option value by approximately $2$--$3\%$ in preliminary tests. This reflects both the seasonal shape and the higher average levels of $\hat\alpha$ and $\sigma_2$, and should be interpreted as the price impact of the seasonal assumptions rather than as a risk premium. The more meaningful question is whether seasonal patterns calibrated to actual futures curves and implied volatilities remain reasonably stable from one crop year to the next (Section~\ref{sec:market}). Note also that \eqref{eq:futures} holds for constant parameters. With a time-dependent $\hat\alpha(\tau)$ and $\sigma_i(\tau)$, the function $A$ is obtained by integrating its ordinary differential equation in time, while $B$ is unchanged as long as $\kappa$ is constant.

\subsection{Greeks and Hedge Ratios}\label{sec:greeks}

In a two-factor model a single spot delta is not sufficient to describe the risk of the position. In the transformed variables the sensitivities are computed by central differences on the grid,
\begin{equation}
\Delta_S=\frac{\partial V}{\partial S}=\frac{V_x}{S},\qquad
\Gamma_S=\frac{\partial^2V}{\partial S^2}=\frac{V_{xx}-V_x}{S^2},\qquad
\Delta_\delta=\frac{\partial V}{\partial\delta}=V_y .
\end{equation}
For an at-the-money option ($S_0=K=100$) the preliminary results are given in Table~\ref{tab:ext}.

\begin{table}[h!]
\centering
\caption{Preliminary extension results (at-the-money put, $S_0=K=100$).}
\label{tab:ext}
\begin{tabular}{lc}
\hline
Quantity & Value \\
\hline
Spot delta $\Delta_S$ & $-0.520$ \\
Convenience-yield delta $\Delta_\delta$ & $16.814$ \\
Price impact of seasonal assumptions & $\approx +2$--$3\%$ \\
\hline
\end{tabular}
\end{table}

The spot delta is close to the value of about $-0.5$ expected for an at-the-money put. The yield delta can be checked against the model structure. Since $\partial F/\partial\delta=-B(T)F$, a put whose value depends on the state mainly through the forward satisfies
\begin{equation}
\Delta_\delta\approx\Delta_F\,\frac{\partial F}{\partial\delta}=|\Delta_F|\,B(T)\,F\approx0.5\times0.352\times98\approx17,
\end{equation}
which is consistent with the computed $16.814$. Because $\delta$ is expressed as a decimal, this corresponds to a change in option value of about $0.17$ for a one-percentage-point move in the convenience yield.\\\\
The convenience yield is not directly tradable, so from a trading perspective the exposure should be expressed as hedge ratios in two futures maturities, for example a nearby contract $F_1$ and a deferred contract $F_2$ with times to delivery $T_1\ne T_2$. Using $\partial F_k/\partial S=F_k/S$ and $\partial F_k/\partial\delta=-B(T_k)F_k$, matching both exposures gives the linear system
\begin{equation}\label{eq:hedge}
\begin{pmatrix}F_1/S & F_2/S\\ -B(T_1)F_1 & -B(T_2)F_2\end{pmatrix}
\begin{pmatrix}h_1\\h_2\end{pmatrix}=\begin{pmatrix}\Delta_S\\\Delta_\delta\end{pmatrix},\qquad
\det=\frac{F_1F_2}{S}\big(B(T_1)-B(T_2)\big),
\end{equation}
where the hedge consists of a position of $-h_k$ in contract $F_k$. The system is solvable when $B(T_1)\neq B(T_2)$, that is $T_1\neq T_2$.

\subsection{Calibration and Market Testing}\label{sec:market}

Rather than adding further theoretical extensions, the most useful next step is a test on a real agricultural option chain, for example December corn or November soybeans.\\\\
This would be done by fitting parameters to the observed futures curve and implied volatilities, and assessing whether the fitted seasonal parameters are stable from one crop year to the next. Further testing from commodity market experts may likely help create a more accurate and usable model for seasonality than proposed in ~\ref{sec:season}.\\\\
Then, we can compare the model prices with Black's (1976) formula and with actual market premiums across strikes.\\\\
Finally, we can test whether the two-futures hedge \eqref{eq:hedge} behaves sensibly around harvest or a sharp weather move, for example by reporting the variance of hedged profit and loss under shocks $(\Delta\ln S,\Delta\delta)$ to the curve. The practical test of the model is whether it produces stable and usable hedges when the futures curve moves sharply.

\newpage

\section{Conclusion}
This project investigated the application of the Alternating Direction Explicit (ADE) method for pricing American commodity options under the Schwartz two-factor model. The theoretical foundations of the ADE method were presented in its double-sweep form, with the Saul'yev alternating scheme as a variant, including the sweep structure of each, their consistency and stability properties, and their computational advantages over traditional implicit finite difference approaches. The method was then extended to the two-dimensional Schwartz framework by incorporating the stochastic convenience yield factor, coordinate transformations, and appropriate boundary treatments based on Fichera theory.\\\\
The resulting numerical scheme was applied to both European and American commodity options. The European option results were validated against the analytical Schwartz solution and Monte Carlo simulations, demonstrating the accuracy of the finite difference implementation. The American option framework was developed through the variational inequality formulation and implemented using a projection-based exercise constraint. Numerical results were benchmarked against a Least Squares Monte Carlo approach across multiple market scenarios, with the ADE method producing consistently accurate results while retaining the computational efficiency of an explicit scheme. The convergence study also showed that the time step must be refined along with the spatial step, and the double-sweep scheme was compared with the Saul'yev variant in terms of accuracy and computational cost.\\\\
Finally, extensions motivated by agricultural commodity markets were tested: an exercise obstacle on a futures contract, seasonal parameters, and Greeks with respect to both state variables, from which hedge ratios in two futures maturities can be obtained. Their stability properties and calibration to market data remain subjects for further work.\\\\
Overall, this study demonstrates that the ADE method provides an effective numerical framework for solving a two-dimensional pricing problem arising from American commodity options with stochastic convenience yields. The combination of the Schwartz two-factor model and ADE finite difference methodology offers a practical approach for accurately capturing commodity-specific dynamics while maintaining computational efficiency.


\newpage
\begin{thebibliography}{99}

\bibitem{barakat1966} Barakat, H. Z., and Clark, J. A. (1966). On the Solution of the Diffusion Equations by Numerical Methods. \textit{Journal of Heat Transfer}, 88(4), 421--427.

\bibitem{black1976} Black, F. (1976). The pricing of commodity contracts. \textit{Journal of Financial Economics}, 3(1-2), 167--179.

\bibitem{blackscholes1973} Black, F., and Scholes, M. (1973). The Pricing of Options and Corporate Liabilities. \textit{Journal of Political Economy}, 81(3), 637--654.

\bibitem{brennan1958} Brennan, M. J. (1958). The Supply of Storage. \textit{The American Economic Review}, 48(1), 50--72.


\bibitem{cranknicolson1947} Crank, J., and Nicolson, P. (1947). A practical method for numerical evaluation of solutions of partial differential equations of the heat-conduction type. \textit{Mathematical Proceedings of the Cambridge Philosophical Society}, 43(1), 50--67.

\bibitem{duffy2004} Duffy, D. J. (2004). Financial Instrument Pricing Using C++. Wiley.

\bibitem{duffy2006} Duffy, D. J. (2006). Finite Difference Methods in Financial Engineering: A Partial Differential Equation Approach. Wiley.

\bibitem{duffy2009} Duffy, D. J. (2009). Unconditionally Stable and Second-Order Accurate Explicit Finite Difference Schemes Using Domain Transformation: Part I One-Factor Equity Problems.

\bibitem{duffy2022} Duffy, D. J. (2022). \textit{Numerical Methods in Computational Finance: A Partial Differential Equation (PDE/FDM) Approach}. John Wiley \& Sons.

\bibitem{fichera1960} Fichera, G. (1960). On a unified theory of boundary value problems for elliptic-parabolic equations of second order. In \textit{Boundary Problems in Differential Equations}. University of Wisconsin Press.

\bibitem{gibsonschwartz1990} Gibson, R., and Schwartz, E. S. (1990). Stochastic Convenience Yield and the Pricing of Oil Contingent Claims. \textit{The Journal of Finance}, 45(3), 959--976.

\bibitem{hilliardreis1998} Hilliard, J. E., and Reis, J. (1998). Valuation of Commodity Futures and Options under Stochastic Convenience Yields, Interest Rates, and Jump-Diffusion. \textit{Journal of Futures Markets}, 18(1), 61--82.

\bibitem{longstaffschwartz2001} Longstaff, F. A., and Schwartz, E. S. (2001). Valuing American Options by Simulation: A Simple Least-Squares Approach. \textit{The Review of Financial Studies}, 14(1), 113--147.

\bibitem{merton1973} Merton, R. C. (1973). Theory of Rational Option Pricing. \textit{The Bell Journal of Economics and Management Science}, 4(1), 141--183.


\bibitem{miltersenschwartz1998} Miltersen, K. R., and Schwartz, E. S. (1998). Pricing of Options on Commodity Futures with Stochastic Term Structures of Convenience Yields and Interest Rates. \textit{Journal of Financial and Quantitative Analysis}, 33(1), 33--59.

\bibitem{peacemanrachford1955} Peaceman, D. W., and Rachford, H. H. (1955). The numerical solution of parabolic and elliptic differential equations. \textit{Journal of the Society for Industrial and Applied Mathematics}, 3(1), 28--41.

\bibitem{pirrong2012} Pirrong, C. (2012). \textit{Commodity Price Dynamics: A Structural Approach}. Cambridge University Press.

\bibitem{rigatos2018} Rigatos, G., Siano, P., and Abbaszadeh, M. (2018). Kalman Filter-based parameter estimation for the Schwartz two-factor commodity pricing model. \textit{IET Cyber-Phys. Syst., Theory Appl.}, Vol. 3 Iss. 4, pp. 232-245

\bibitem{saulyev1964} Saul'yev, V. K. (1964). \textit{Integration of Equations of Parabolic Type by the Method of Nets}. Pergamon Press Oxford.

\bibitem{schwartz1997} Schwartz, E. S. (1997). The Stochastic Behavior of Commodity Prices: Implications for Valuation and Hedging. \textit{The Journal of Finance}, 52(3), 923--973.

\bibitem{working1949} Working, H. (1949). The Theory of the Price of Storage. \textit{The American Economic Review}, 39(6), 1254--1262.

\end{thebibliography}
\end{document}